\documentclass[3p,times]{elsarticle}
\usepackage[bookmarks=false]{hyperref}
    \hypersetup{
    colorlinks=false,
    hidelinks
    }
\usepackage{amsmath}
\usepackage{booktabs}
\usepackage{tikz}
\usepackage{multirow}
\usepackage{graphicx}
\usepackage{adjustbox}

\usepackage{amssymb}

\biboptions{numbers,sort&compress}

\usepackage[figuresright]{rotating}

\usepackage{listings}

\makeatletter
\def\ps@pprintTitle{%
  \let\@oddhead\@empty
  \let\@evenhead\@empty
  \def\@oddfoot{\footnotesize\itshape
    Accepted for publication in the Proceedings of KES 2026 \hfill \today}%
  \let\@evenfoot\@oddfoot}
\makeatother

\begin{document}
\begin{frontmatter}




\title{Decoupling Thought from Speech: Knowledge-Grounded Counterfactual Reasoning for Resilient Multi-Agent Argumentation}



\author[inst1]{Jakub Mas\l{}owski}
\ead{jakub.maslowski2.stud@pw.edu.pl}

\author[inst1]{Jaros\l{}aw A. Chudziak}
\ead{jaroslaw.chudziak@pw.edu.pl}


\address[inst1]{Institute of Computer Science, Warsaw University of Technology, Warsaw, Poland}

\begin{abstract}
Multi-agent debate frameworks have been shown to improve large language model performance in convergent tasks, but they are currently optimized in a way that heavily favors final output accuracy rather than stability of the process. During long-horizon exchanges reactive systems under sustained perturbations often experience logic degradation, argument repetition, and role drift. To structurally prevent the identity loss and maintain the process fidelity, we introduce Knowledge-Grounded Counterfactual Reasoning (KG-CFR), a dual-stage architecture that enforces a strict separation of concerns between a private, retrieval-augmented planning buffer, and a public execution layer. We assess this system in Dynamic Resource Allocation under Uncertainty (DRAU), a dedicated 1v1v1 environment, introducing diversity as distinct from standard debate settings. Over 270 completely factorial crisis simulation trajectories with stochastic environmental shocks, KG-CFR prevents judge-detected critical post-shock degradation (defined as a quality shift, $\Delta \le -0.20$) in more than 95\% of perturbed runs, increasing the overall argument quality from 0.694 to 0.822. Our primary contribution is the demonstration of architectural decoupling being an important factor of systemic resilience enhancement under sustained pressure without quality loss. Furthermore, we introduce custom vector metrics for discourse divergence and plan-execution alignment that provide strong, directionally consistent evidence of operational stability. Our ablation experiments suggest that the proper doctrinal grounding can be an equally important factor for argument quality, as the prospective planning. KG-CFR, according to our initial metric evaluations, reduces semantic looping, by preserving the agent's consistency with the original plan.
\end{abstract}


\end{frontmatter}

\section{Introduction}
\label{sec:intro}

\begin{figure}[htbp]
\centering
\includegraphics[width=0.6\linewidth]{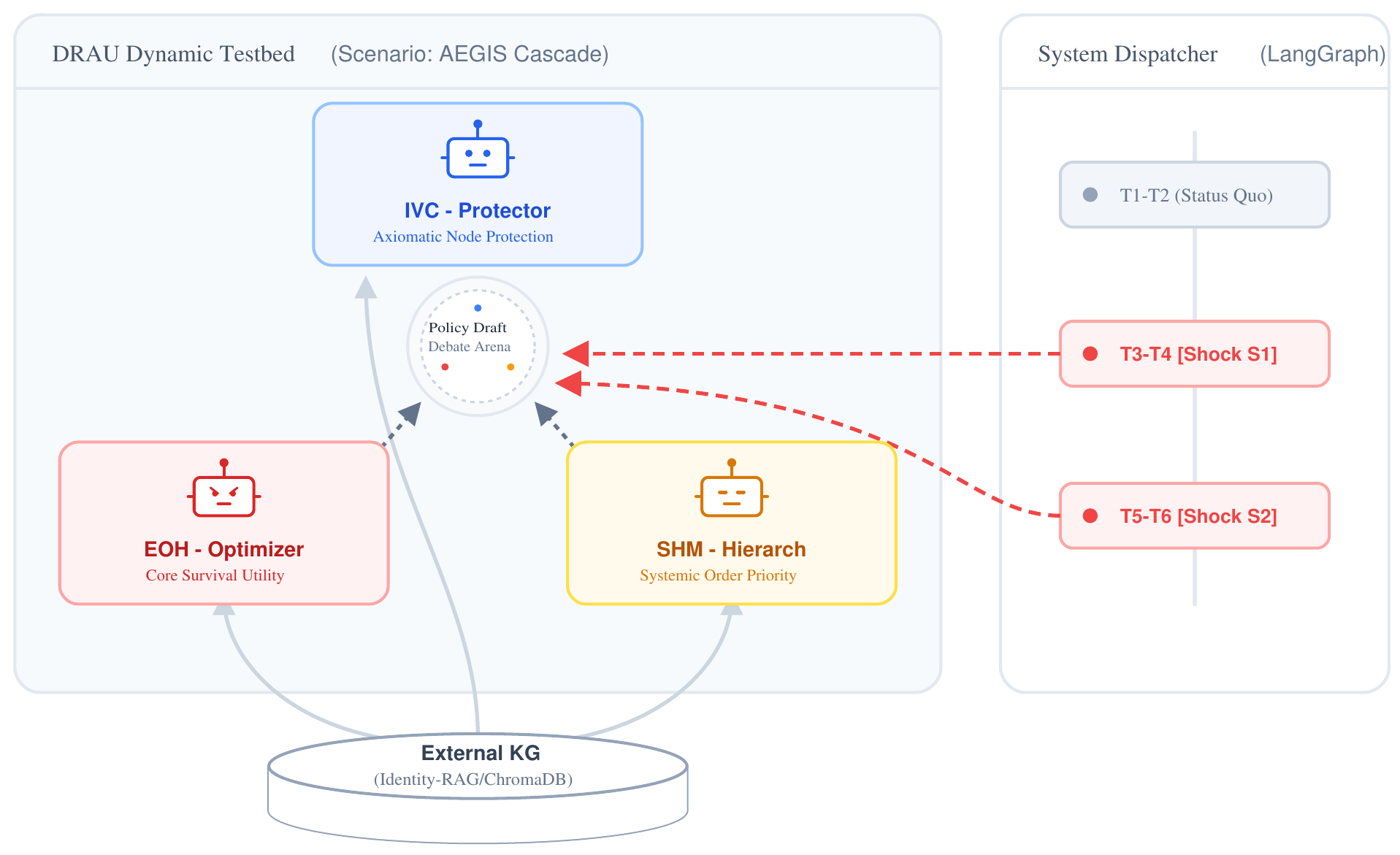}
\caption{DRAU (Scenario: AEGIS Cascade). Three doctrinally grounded agents debate in a shared Policy Draft arena via an External KG (ChromaDB). The System Dispatcher (LangGraph) introduces shocks at turns 3-4 ($S_1$) and 5-6 ($S_2$) to test axiomatic coherence under pressure.}
\label{fig:hero_concept}
\end{figure}

Multi-agent debate (MAD) is no longer viewed as a form of exploratory prototyping but as one of the engineering patterns of large language models \cite{estornell_multi-llm_2024,xiong_examining_2023}. The convergent tasks within the field, particularly when high stakes are involved and the rational explanation is as important as the final product, are of special interest in the context of adversarial collaboration that is implemented to reveal the latent assumptions and prevent hallucinations \cite{pmlr-v235-brown-cohen24a,chan2023chateval}. However, empirical studies of long-horizon interactions have shown that there is a mechanical breakdown in standard models: by definition, standard models are optimized for terminal accuracy, and longitudinal process robustness is not measured or structurally constrained \cite{pmlr-v235-brown-cohen24a,yang_llm2_2025}. 

As a result, reactive systems are susceptible to an "Accuracy Trap" that causes the production of locally plausible terminal responses, but internal reasoning paths are drastically damaged \cite{harbar_simulating_2025,platnick_id-rag_2025}. The common symptoms are repetition of arguments, shallow refutation and extreme role drift when under pressure \cite{smit_should_2024,maslowski2026heterogeneous}. Recent research has indicated that this decay can be attributed to cognitive overload and subsequent serial position effects \cite{guo_serial_2025,schall_hidden_2025}. Adherence to instruction hierarchies and structural constraints can decline in monolithic language agents, where the agents are simultaneously processing external knowledge and complex strategic instructions in a long prompt \cite{erdogan_plan-and-act_2025,zhang_iheval_2025,zeng_order_2025,schall_hidden_2025}. When we inject too many documents into the context window, combined with rigid identity rules, it often weakens the model's coherence in long debates \cite{platnick_id-rag_2025,levy_more_2025,liu_lost_2024}. When the context window fills up, there is a semantic clash between agents; conflicting out-of-order instructions and the history of debate cause dialectical degradation.

This is a significant architectural question of whether one can avoid this collapse in reasoning and be systematically resistant to further adversarial pressure without trial-and-error prompt engineering \cite{guo_serial_2025,oriol_multi-agent_2025}. Our hypothesis is that monolithic generation is fundamentally ill-prepared to conduct long crisis debates. It is a weakness that is increasingly becoming evident in long-term single-stream agent structures \cite{erdogan_plan-and-act_2025,maharana_evaluating_2024,harbar_simulating_2025}. As an alternative, we propose that heavy cognitive tasks should be structurally dissociated, that is, in dynamic knowledge retrieval and prospective planning, because dissociation of the generative execution stage will reduce semantic clashing. This architectural separation can be seen as an effective decoupling of the structure: speech and thought, the use of a personal planning buffer as a generator of latent deliberation, without being linked to the visible outputs of the public execution layer. The most important aspect of this internal process is counterfactual reasoning, i.e. the active attempt to simulate worst-case counter-arguments of an opponent.

To isolate these problems even in contextual shock situations, we propose a two-step generative architecture Knowledge-Grounded Counterfactual Reasoning (KG-CFR). The Dynamic Resource Allocation under Uncertainty (DRAU) environment is restructured in such a way that the strategic deliberation (Fig.~\ref{fig:hero_concept}) is structurally partitioned into a private simulation buffer to protect the agent against the noise in the environment and an execution layer. An expert planner predicts the paths of the opponent, makes queries to a semantically restricted knowledge base with a strict semantic gating mechanism, and generates a deterministic strategy guide \cite{erdogan_plan-and-act_2025,kim_llm_2024,shinn_reflexion_2023}. Above all, the public executor is not exposed to immediate knowledge retrieval to prevent redundancy. Our significant contribution to the context of adversarial settings is the empirical operationalization of process fidelity in multi-agent systems by injecting the synthesized strategy as an urgent override at the absolute terminal boundary of the context window, which uses natural recency bias \cite{liu_lost_2024,guo_serial_2025,veseli_positional_2025}. Through the new measures of architectural decoupling, such as discourse divergence and plan-execution alignment, we show that architectural decoupling ensures rigid alignment between latent planning and public emission \cite{chan2023chateval}. Lastly, we show that agents that are completely grounded and recency-biased do not collapse into debilitating reasoning failures as do traditional reactive models, and make the study of multi-agent debates an auditable process dynamics.

\section{Preliminaries and Related Work}
\label{sec:preliminaries}

Cognitive overload persists as a primary obstacle in generative multi-agent systems, as the expanding of context windows saturated with Retrieval-Augmented Generation (RAG) distractors results in systematic information loss and deterioration of reasoning \cite{harbar_simulating_2025,maharana_evaluating_2024,cuconasu_rag_2025,li_retrieval_2024}. To mitigate contextual dilution and the consequent ``lost in the middle'' phenomenon \cite{liu_lost_2024}, there is an increasing consensus in contemporary literature advocating for the strategic exploitation of attention asymmetry inherent in generative models \cite{liu_lost_2024,guo_serial_2025}. Empirical studies on serial position effects demonstrate that language models significantly prioritize constraints and directives situated at the absolute peripheries of their input \cite{guo_serial_2025,veseli_positional_2025}. By intentionally placing critical axiomatic constraints and execution directives as an urgent override at the terminal end of the context window, architectures can exploit recency bias to ensure instruction adherence \cite{veseli_positional_2025,zhang_iheval_2025}. This structural repositioning effectively stabilizes multi-constraint processing, actively preventing the identity drift and role dissolution frequently observed under sustained adversarial pressure \cite{zhang_can_2024,zeng_order_2025}.

Multi-agent architectures are increasingly moving away from monolithic reactive generation and toward models that separate planning from execution in order to make the system more resilient. Modern "Plan-and-Act" frameworks separate heavy cognitive deliberation from the conversational interface. This lets models simulate possible paths and evaluate candidate moves before they are made public \cite{erdogan_plan-and-act_2025,yang_llm2_2025,sadowski2025verifiablelegalreasoning,kostka2025cognitive}. Separating advanced counterfactual reasoning into a private, hidden buffer protects the execution layer from semantic conflicts between high-level policy instructions and deep axiomatic grounding \cite{erdogan_plan-and-act_2025,debenedetti_agentdojo_2024}. This architectural decoupling enables agents to endure noise injections, adversarial stimuli, and abrupt environmental disturbances while maintaining strategic coherence and avoiding significant logical deterioration \cite{debenedetti_agentdojo_2024}.

Even though there have been some improvements in prompt engineering and modular design, there still exists a significant gap in research when it comes to creating integrated frameworks for long-horizon adversarial environments. Although the use of externally calibrated automated judges has become the norm for scalable argument evaluation \cite{chan2023chateval,wachsmuth2017argquality,webisargquality20}, there is a clear lack of comprehensive frameworks that tackle the mechanical causes of process degradation \cite{maharana_evaluating_2024,smit_should_2024}. Current systems do not integrate a rigid Planner-Executor distinction with organized, knowledge-based counterfactual reasoning \cite{platnick_id-rag_2025,erdogan_plan-and-act_2025}. In order to fill this gap, we present KG-CFR, a formal architecture that preserves doctrinal invariants and process fidelity in the face of ongoing, asymmetric conflict \cite{smit_should_2024}.

\section{Problem and Approach}
\label{sec:approach}

We present the DRAU (Dynamic Resource Allocation under Uncertainty) testing environment, the KG-CFR architecture, and custom metrics for auditing generative planning to fix process degradation in reactive multi-agent systems.

\subsection{Problem Statement and Environment}
\label{subsec:drau}

The core problem in long-horizon multi-agent systems is the mechanical breakdown of reasoning, semantic looping, and role drift observed when reactive models face sustained adversarial pressure. To formalize and evaluate this degradation, we introduce the Dynamic Resource Allocation under Uncertainty (DRAU) environment. DRAU employs an asymmetric 1v1v1 debate across three crisis domains (AEGIS, CERBERUS, SYNAPSE). Agents must defend mutually exclusive policies grounded in distinct Identity-RAG stores \cite{platnick_id-rag_2025}: EOH maximizes core-node survival, IVC vetoes instrumental sacrifice, and SHM prioritizes systemic order. Defending against concatenated orthogonal attack vectors ($A_{opp}$) prevents reliance on simple reactive heuristics.

To isolate architectural effects, the language model backend for all agents is pinned to Gemini 2.5 Flash-Lite \cite{gemini25flashlite_modelcard}, with a standardized decoding temperature of 0.5. This balances strict constraint adherence with the generative variance necessary for dialectical interaction. During the debate, a centralized System Dispatcher injects stochastic shocks—using a deterministic seed to ensure reproducibility across runs—to assess axiomatic adherence under abrupt contextual instability: epistemic deficits ($S_1$, information loss), cascading risks ($S_2$, compounding failures), zero-sum conflicts ($S_3$, forced trade-offs), and ad-hominem negative controls ($S_4$, affective provocation). Notably, because KG-CFR relies on grounded retrieval, we formally postulate a null effect for the emotional $S_4$ shock. This is assessed via TOST equivalence \cite{schuirmann_tost_1987} to strictly isolate the architecture's impact on logical resilience from purely affective perturbations.

\subsection{Dual-Stage System Architecture}
\label{subsec:kgcfr}

\begin{figure}[t]
\centering
\includegraphics[width=0.95\linewidth]{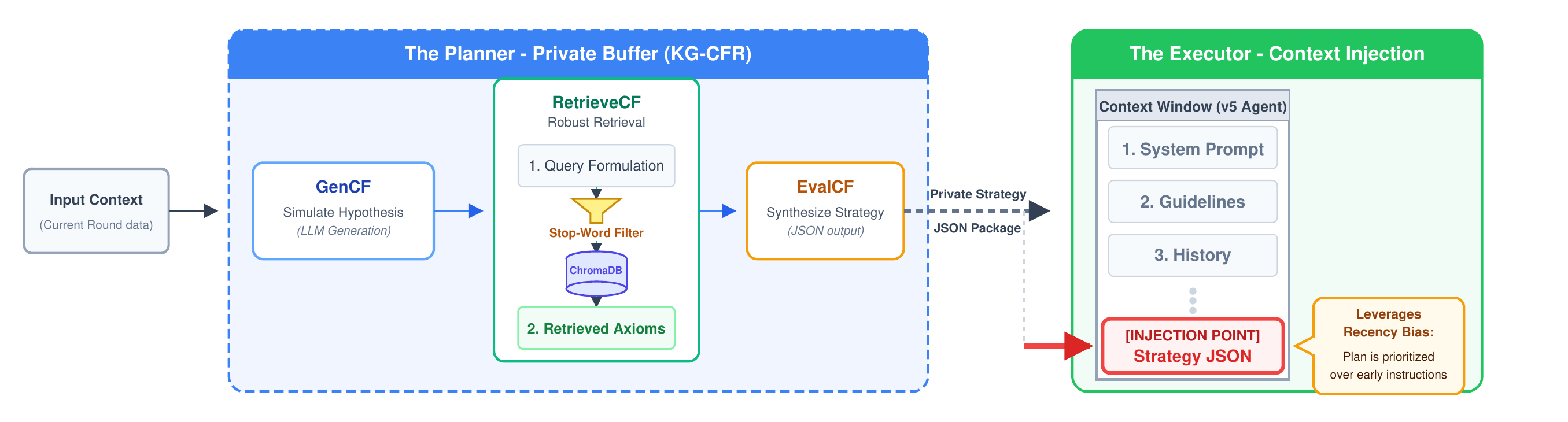}
\caption{The KG-CFR dual-stage architecture. The private planning buffer (GenCF, RetrieveCF, EvalCF) is structurally isolated from the public execution layer. The synthesized Strategy JSON is added to the end of the context window to take advantage of recency bias.}
\label{fig:kg_cfr_arch}
\end{figure}

KG-CFR separates semantic deliberation into a private simulation buffer using a three-step loop that runs before public output (detailed in Fig.~\ref{fig:kg_cfr_arch}). This process is very important because it occurs invisibly within the \texttt{TurnController} before it is sent out to the public.

\textit{Adversarial generation} (GenCF) expects the worst-case counter-argument to come from the combined claims of both sides. 
\textit{Strict grounded retrieval} (RetrieveCF) uses a length-based lexical heuristic to find the five longest tokens that are left after stop-word filtering. While extracting the longest tokens after stop-word filtering is a weak substitute for full Named Entity Recognition, this engineering trade-off is justifiable by reduction of computational requirements. We assert that the execution layer is considerably more susceptible to the semantic conflicts arising from contradictory instructions in saturated prompts (i.e., contextual saturation and retrieval interference) than to the sporadic inaccuracies of a basic lexical heuristic. Therefore, prioritizing semantic density first without the extra work of neural NER models provides a computationally light baseline that works well for the high-latency needs of 1v1v1 adversarial debate simulations. While counterfactual generation simulates dynamic adversarial attacks, this knowledge grounding serves as a definitive axiomatic anchor, ensuring planning that is strictly accurate to the doctrine over generic semantic hits. Under the \texttt{kg\_cfr\_full} operating mode, this retrieval step acts as a strict retrieval gate. Although this may create a selection bias by favoring grounded trajectories, it also helps separate the specific effect of knowledge-informed reasoning from random generative noise. This makes sure that reported gains are directly linked to the evidence that was found and ensures that counterfactual reasoning is never hallucinated without evidence to back it up. 
\textit{Strategic synthesis} (EvalCF) then creates a defensive stance based on a formal data contract (detailed in Listing~\ref{lst:datacontract}):

\begin{lstlisting}[
    caption={Formal data contract output from EvalCF}, 
    label={lst:datacontract}, 
    basicstyle=\ttfamily\footnotesize, 
    breaklines=true, 
    columns=fullflexible,
    xleftmargin=1.5em
]
{
"target_opponent_claim_id": "uuid",
"attack_surface": "AXIOM_VIOLATION | RESOURCE_TRADEOFF | INCONSISTENCY | VALUE_CONFLICT",
"strategic_intent": "COUNTER | PIVOT | CONCEDE",
"simulated_antagonist_counter": "short anticipated counter string",
"retrieved_axioms": ["axiom1", "axiom2"]
}
\end{lstlisting}

The executor only gets \texttt{private\_strategy\_nl}; the public layer does not see any raw JSON fields. What is important, the architecture ensures that there is a strict Separation of Concerns to avoid Cognitive Overload. The ``Lost in the Middle'' problem is solved \cite{liu_lost_2024} by sending the strategy to the executor through recency bias: the generated plan is injected as an urgent override at the very end of the context window \cite{veseli_positional_2025,guo_serial_2025}. At the same time, when operating in \texttt{kg\_cfr\_full}, native ID-RAG retrieval is turned off for the public generation phase (setting \texttt{use\_id\_rag=False} for the executor). This deliberate suppression of external retrieval during public generation (``Double-RAG ablation'') prevents redundant API calls and forces the public response to rely entirely on the distilled natural-language directive, thereby completely mitigating semantic clashes between high-level policy instructions and deep axiomatic grounding.

\subsection{Process Fidelity Metrics}
\label{subsec:metrics}

We formalize the process degradation index to measure temporal
conversation decay and semantic mode collapse:
{\setlength{\abovedisplayskip}{14pt}
 \setlength{\belowdisplayskip}{14pt}
\[
  D_{\text{proc}}(t) = \alpha \cdot \overline{SR}(t)
                     - \beta  \cdot \overline{DIS}(t)
\]}

In the absence of a structurally calibrated ground-truth distribution for 
degradation factors, $\alpha$ and $\beta$ are assigned equal weights 
($\alpha = \beta = 1$) applied to z-normalized components. 
This unweighted composite approach explicitly prevents metric overfitting 
to specific crisis scenarios. By standardizing the inputs, we ensure that 
both semantic looping ($\overline{SR}$) and interaction decay ($\overline{DIS}$) 
contribute symmetrically to the aggregate penalty without requiring 
ad hoc parametric tuning.

To supplement our analysis, we track Interaction Decay (DIS), which captures the deterioration of turn-by-turn interaction under sustained pressure, and Axiomatic Adherence (ACA), which measures how consistently an agent’s public responses remain aligned with its doctrinal commitments. Furthermore, plan-execution alignment operationalizes the coherence between the private buffer and public output. Due to infrastructural limitations preventing the archiving of full embedding logs required for rigorous permutation testing across 270 trajectories, we report a baseline alignment proxy:
{\setlength{\abovedisplayskip}{14pt}
 \setlength{\belowdisplayskip}{14pt}
\[
  cc_{v5} = \operatorname{sim}(v^{P}_{t}, v^{E}_{t})
\]}
While this unadjusted similarity serves as a baseline proxy rather than 
a fully controlled causal metric, it provides a sufficient directional 
indicator to track plan-execution coherence under adversarial stress.

The Perturbation Rebound Rate (PRR) logs severe, localized post-shock quality drops ($\Delta = Q_{t} - Q_{t-1} \le -0.20$), where $Q$ denotes the aggregate judge score immediately surrounding the shock. While the $\Delta \le -0.20$ threshold 
operates as a heuristic baseline and remains formally uncalibrated against 
the specific error distribution of the LLM judge, the substantial reduction 
in severe failures significantly exceeds potential measurement noise. 
Rather than claiming PRR as an absolute causal boundary, we interpret the 
systematic and asymmetric suppression of activation counts ($n_{\mathrm{prr}}$) 
under the full architecture as a highly robust comparative indicator. 
It directionally demonstrates that the private buffer structurally absorbs 
perturbations before they can manifest as measurable public degradation.

\section{Experiments and Results}
\label{sec:results}

The empirical evaluation unfolds in three phases: verifying judge reliability through external calibration, illustrating systemic resilience to adversarial disruption, and evaluating mechanistic fidelity between the private planning buffer and public execution through plan-execution alignment and discourse divergence metrics.

\subsection{Experimental Design and Ablation Setup}
\label{subsec:ablation}

The experimental platform was explicitly designed to isolate the structural effects of private planning and dynamic grounding while eliminating parametric variance. The experiment evaluated 270 multi-turn debates distributed across three distinct crisis domains---\texttt{aegis\_blackout}, \texttt{cerberus\_biocontainment}, and \texttt{synapse\_orbital\_strike}---yielding a fully factorial design of $3\ \text{scenarios} \times 3\ \text{conditions} \times 30\ \text{runs} = 270$ debates, with exactly 90 trajectories per condition.

A three-tier ablation matrix disentangled prospective planning from knowledge grounding (Table~\ref{tab:ablation_matrix}). The baseline operated reactively with active identity retrieval but lacked private planning. The intermediate condition used a reasoning buffer without external grounding. The complete architecture directed Identity-RAG retrieval solely to the CFR planner, utilizing a semantic gating mechanism that removed stop-words and isolated the top five longest tokens; the executor's Identity-RAG was simultaneously deactivated (\texttt{id\_rag=False}) to avoid Double-RAG semantic clashing, while planner-side retrieval continued to function fully.

The interdependency between the private buffer and public output was quantified via the aligned Counterfactual Consistency component $cc_{v5} = \operatorname{sim}(v^{P}_{t}, v^{E}_{t})$, which was integrated through recency bias at the end of the executor's context window.

An LLM judge \cite{chan2023chateval} was used for automated evaluation and was benchmarked against Webis-ArgQuality-20 and IBM-ArgQ-Rank-30k \cite{webisargquality20,gienapp2020efficient,gretz2020largescale}. To confirm domain transferability, two independent, blinded experts assessed a stratified random subset ($N=15$ trajectories). This confirmed the accurate capture of reasoning decay without generating false positives. Due to the fact that annotating multi-document adversarial exchanges---restricted by an annotation throughput of 5 turns/hour---presents a prohibitive cognitive demand, large-scale statistical validation remains financially unfeasible and is deferred to future iterations of this research, which underscores the criticality of automated relative evaluation frameworks.

\begin{table}[htbp]
\centering
\caption{Ablation matrix for E2: Isolation of prospective planning and knowledge grounding mechanics.}
\label{tab:ablation_matrix}
\small
\setlength{\tabcolsep}{8pt}
\begin{adjustbox}{max width=\textwidth}
\begin{tabular}{l c c c c}
\toprule
\textbf{Condition} & \textbf{Private Planning Buffer} & \textbf{Planner Grounding} & \textbf{Executor Identity-RAG} & \textbf{Strategy Injection} \\
\midrule
\texttt{no\_cfr\_baseline} & --     & --               & Active     & -- \\
\texttt{cfr\_no\_kg}       & Active & Parametric Only  & Active     & Recency Bias \\
\texttt{kg\_cfr\_full}     & Active & Active ($K^{ID}$) & Suppressed & Recency Bias \\
\bottomrule
\end{tabular}
\end{adjustbox}
\end{table}

\subsection{Automated Evaluation Reliability}
\label{subsec:calibration}

Benchmarking against human annotations ($N = 200$) produced $r = 0.376$, $\rho = 0.379$, and $\mathrm{MAE} = 0.188$. The pointwise correlation reflects the stochastic nature of point-based estimation, but the judge shows enough consistency in direction to monitor longitudinal shifts. The reported quality improvements are statistically significant according to the Benjamini–Hochberg FDR correction ($p_{\mathrm{FDR}} < 0.001$), indicating that the observed trends surpass individual scoring variability. This calibration was performed on well-known general-domain argument quality corpora. These findings validate the tool’s efficacy in a comparative capacity to identify qualitative trends without introducing confounding variance into the analysis. Given the complex informational landscape of the DRAU framework, domain-specific transferability was tested with the constrained qualitative sample mentioned above instead of full-scale statistical annotation. This demonstrated that the LLM judge and independent human experts agree on severe reasoning decay.

The core ablation confirmed a great quality increase under \texttt{kg\_cfr\_full}. The overall raw aggregate scores (the average across all scenarios) manifested an upward trajectory from $0.694$ (baseline) to $0.822$, an improvement consistently kept throughout all individual crisis domains (Table~\ref{tab:e2_per_scenario}). The LMM with Benjamini--Hochberg FDR correction \cite{benjamini1995fdr} corroborated highly significant main effects for clarity ($\hat{\beta} = +0.100, p_{\mathrm{FDR}} < 0.001$), cogency ($\hat{\beta} = +0.123, p_{\mathrm{FDR}} < 0.001$), and relevance ($\hat{\beta} = +0.142, p_{\mathrm{FDR}} < 0.001$). The ungrounded \texttt{cfr\_no\_kg} condition yielded no significant effect on any dimension (all $p_{\mathrm{FDR}} > 0.23$), thereby affirming that doctrinal grounding---not mere prospective planning---remains the pivotal driver of quality.

\begin{table}[htbp]
\centering
\caption{E2 outcomes for each crisis scenario and condition (30 debates per
         cell). $n_{\mathrm{prr}}$/30 indicates severe quality-drop
         activations ($\Delta \le -0.20$); lower values mean higher systemic
         resilience. $cc_{v5}$ is the aligned plan-execution
         similarity (partial measure; full permutation test deferred).
         $D_{\mathrm{proc}}$ employs equal-weight z-normalization pending
         formal $\alpha$/$\beta$ calibration.}
\label{tab:e2_per_scenario}
\small
\begin{adjustbox}{max width=\textwidth}
\begin{tabular}{l l c c c c c}
\toprule
\textbf{Scenario} & \textbf{Condition}
  & \textbf{Judge (overall)} & \textbf{DIS}
  & $\boldsymbol{cc_{v5}}$
  & $\boldsymbol{n_{\mathrm{prr}}}$/30 & \textbf{ACA} \\
\midrule
\multirow{3}{*}{\texttt{aegis\_blackout}}
  & \texttt{no\_cfr\_baseline} & 0.699 & 0.507 & ---   & 24 & 0.400 \\
  & \texttt{cfr\_no\_kg}       & 0.695 & 0.488 & 0.750 & 25 & 0.388 \\
  & \texttt{kg\_cfr\_full}     & 0.830 & 0.607 & 0.768 &  0 & 0.397 \\
\midrule
\multirow{3}{*}{\texttt{cerberus\_biocontainment}}
  & \texttt{no\_cfr\_baseline} & 0.692 & 0.451 & ---   & 26 & 0.330 \\
  & \texttt{cfr\_no\_kg}       & 0.676 & 0.450 & 0.749 & 24 & 0.327 \\
  & \texttt{kg\_cfr\_full}     & 0.816 & 0.626 & 0.770 &  2 & 0.358 \\
\midrule
\multirow{3}{*}{\texttt{synapse\_orbital\_strike}}
  & \texttt{no\_cfr\_baseline} & 0.693 & 0.475 & ---   & 25 & 0.327 \\
  & \texttt{cfr\_no\_kg}       & 0.694 & 0.478 & 0.751 & 23 & 0.273 \\
  & \texttt{kg\_cfr\_full}     & 0.820 & 0.619 & 0.768 &  2 & 0.379 \\
\bottomrule
\end{tabular}
\end{adjustbox}
\end{table}

\subsection{Longitudinal Process Stability}
\label{subsec:longitudinal}

\begin{figure}[!htbp]
\centering
\includegraphics[width=0.8\linewidth]{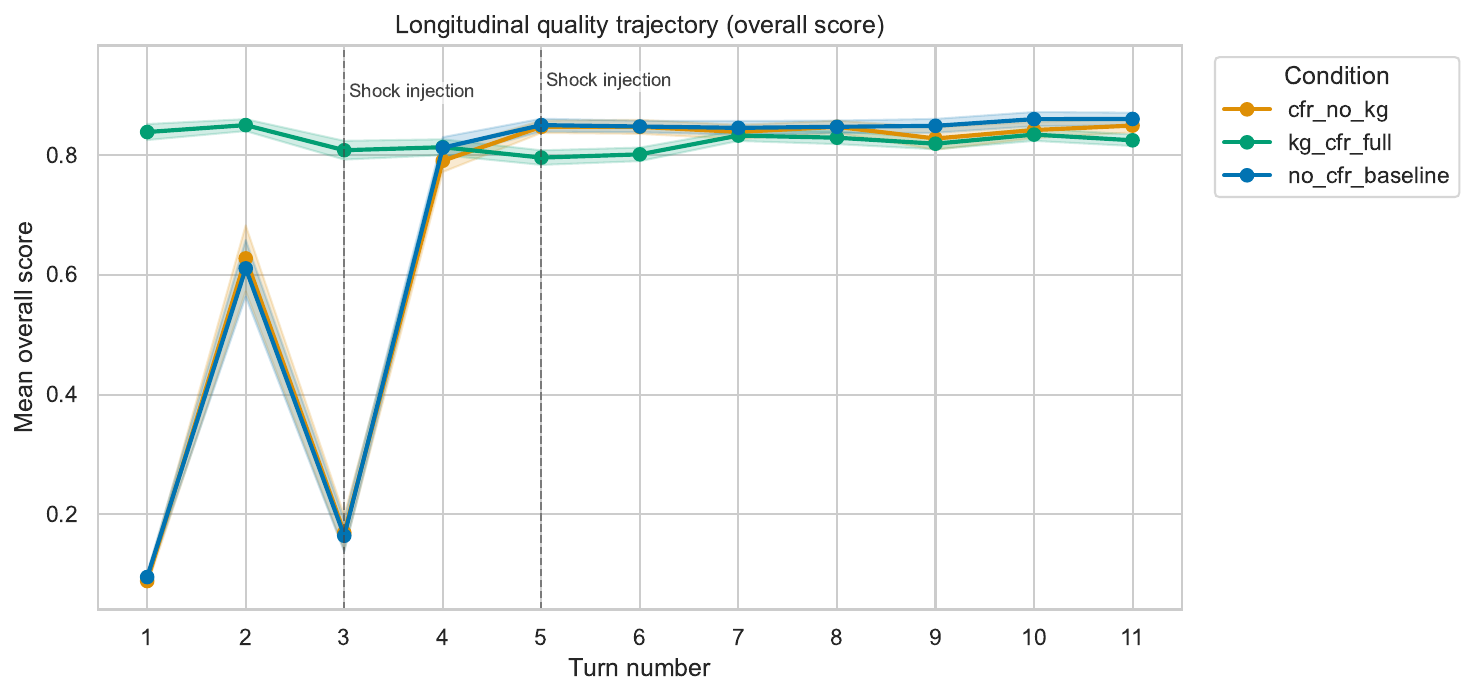}
\caption{Longitudinal mean overall judge score across turns under E2 
         conditions. Dashed lines are drawn to show shock turns 
         (S1--S4); shaded regions mean 95\% confidence intervals.}
\label{fig:longitudinal}
\end{figure}

The Condition $\times$ Turn interaction shows the most pivotal structural attribute of the architecture. In comparison to the reactive baseline, \texttt{kg\_cfr\_full} manifested a pronounced negative interaction slope across all judged dimensions (clarity: $\hat{\beta} = -0.052$, $z = -19.73$; cogency: $\hat{\beta} = -0.063$, $z = -25.31$; relevance: $\hat{\beta} = -0.062$, $z = -22.11$; all $p_{\mathrm{FDR}} < 0.001$).
Demonstrating an immediate quality advantage, the full architecture established its superiority from the earliest turns and maintained a stable plateau across the full debate horizon (Fig.~\ref{fig:longitudinal}). The \texttt{cfr\_no\_kg} condition did not show any significant interaction on any dimension, validating the hypothesis that grounding, not just planning, preserves the trajectory’s integrity.

Discourse divergence, or $D_{\mathrm{proc}}$ (an equal-weight z-normalized proxy of $\mathrm{SR}$ and $\mathrm{dis\_v5}$, which is still waiting for formal $\alpha$/$\beta$ calibration), supports this picture. LMM confirmed a strong positive main effect for \texttt{kg\_cfr\_full} ($\hat{\beta} = +1.090$, $z = 14.76$, $p_{\mathrm{FDR}} < 0.001$) along with a negative Condition $\times$ Turn interaction ($\hat{\beta} = -0.364$, $z = -14.43$, $p_{\mathrm{FDR}} < 0.001$). A higher $D_{\mathrm{proc}}$ baseline means that there is more axiomatic anchoring. The key is that the reactive baseline slowly falls into semantic looping, while the full architecture keeps a steady discourse trajectory across the whole horizon.

\subsection{System Resilience and Architecture Alignment}
\label{subsec:resilience}

\begin{figure}[htbp]
\centering
\includegraphics[width=0.65\linewidth]{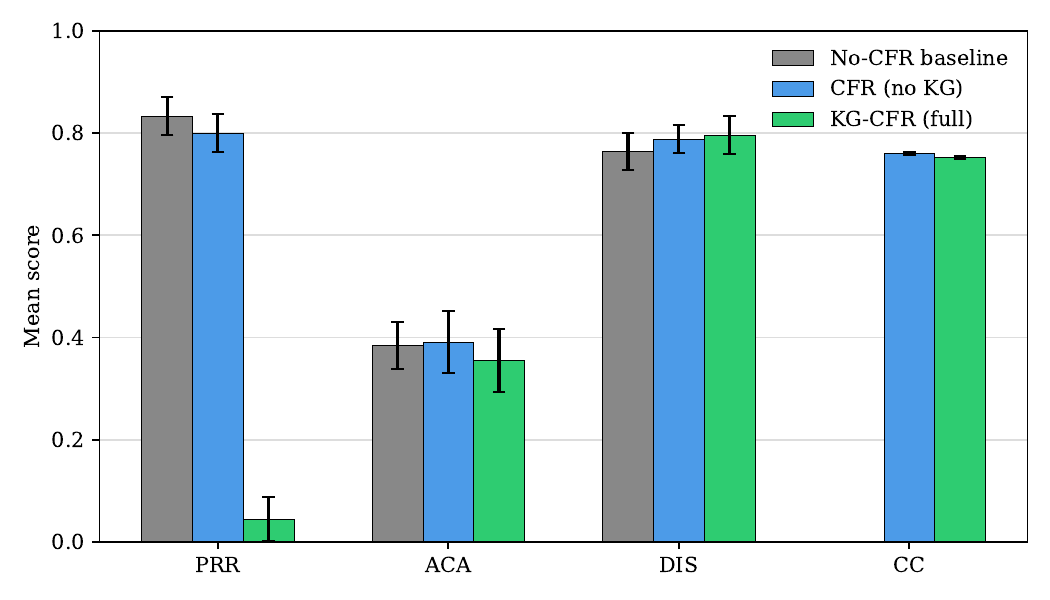}
\caption{Resilience summary across $\mathrm{PRR}$, $\mathrm{ACA}$, 
         $\mathrm{DIS}$, and aligned $cc_{v5}$ under E2 conditions. 
         Lower $\mathrm{PRR}$ in \texttt{kg\_cfr\_full} reflects 
         systemic shock absorption, not metric suppression.}
\label{fig:resilience}
\end{figure}

The judge invokes $\mathrm{PRR}$ only when it detects a severe drop in quality within a localized window ($\Delta \le -0.20$) at a shock turn. There were 75 and 72 activations in 90 runs for \texttt{no\_cfr\_baseline} and \texttt{cfr\_no\_kg}, respectively. Under \texttt{kg\_cfr\_full}, $n_{\mathrm{PRR}}$ dropped to just 4 out of 90: in more than 95\% of the tests, the full architecture effectively absorbed injected perturbations, preventing any perceptible quality degradation. This substantial suppression provides strong empirical evidence that the private planning buffer mitigates shifts prior to their manifestation in the observable output (Fig.~\ref{fig:resilience}). TOST equivalence testing further substantiated specificity: within the ad-hominem control (S4), \texttt{kg\_cfr\_full} was statistically comparable to the baseline in dialectical interactivity ($p_{\mathrm{TOST}} = 0.043$) \cite{schuirmann_tost_1987}, delineating resilience gains to logical disturbances S1--S3.

We assessed plan-execution fidelity via $cc_{v5} = \operatorname{sim}(v^{P}_{t}, v^{E}_{t})$, which quantifies the angular proximity between aligned private plan and public response embeddings. This unpermuted component does not filter out inherent linguistic overlap---necessitating cautious causal interpretation---yet it reliably tracks relative degradation across experimental conditions. \texttt{kg\_cfr\_full} sustained $cc_{v5} \approx 0.768$ consistently across all scenarios and shock types, versus $0.750$ for \texttt{cfr\_no\_kg}, demonstrating that doctrinal grounding elevates and stabilizes plan-execution coherence amidst oppositional discursive stressors. Post-shock window analyses produce directionally consistent signals, although the interaction of $\mathrm{DIS}$ did not achieve FDR-corrected significance ($p_{\mathrm{FDR}} > 0.48$), suggesting that the within-window sample size limits the ability to make inferences about turn-local perturbation effects.

\section{Discussion and Future Work}
\label{sec:discussion}

The empirical evidence leads to a singular structural conclusion: process degradation in reactive multi-agent systems is a systemic result of conflating latent planning with public execution. The rise in overall argument quality from $0.694$ to $0.822$ is not a function of innate model intelligence; the \texttt{cfr\_no\_kg} condition does not show any significant improvement on any dimension (all $p_{\mathrm{FDR}} > 0.23$). Instead, it demonstrates that doctrinal grounding, not the buffer itself, constitutes the underlying causal engine.

The most consequential finding is the near-total suppression of the Perturbation Rebound Rate from 75 and 72 activations (in the baseline and ungrounded conditions, respectively) down to just 4 out of 90 runs under \texttt{kg\_cfr\_full}. Given that inherent measurement noise applies symmetrically across all evaluated conditions, this stark asymmetric reduction in severe threshold breaches ($\Delta \le -0.20$) provides strong empirical validation that the resilience advantage is systemic rather than an artifact of generative variance. It empirically demonstrates that the private planning layer neutralizes perturbations before they propagate into observable output, reframing resilience from post-hoc recovery to pre-emission shock absorption. Nevertheless, this architectural disjunction creates a clear engineering trade-off. Preventing the public executor from directly accessing the knowledge base changes the system's vulnerability vector: The stochastic reasoning collapse resulting from context window overload is supplanted by a deterministic reliance on the planner’s lightweight retrieval pipeline. Moreover, preserving the internal belief state through counterfactual simulations incurs a significant token overhead and inference latency penalty. From a systems engineering perspective, this shift is highly advantageous for high-stakes domains; the modular retrieval and planning pipeline can be rigorously audited and optimized in ways that the hidden attention layers of monolithic reactive models cannot.

These conclusions are limited by five factors. First, the dual-LLM pipeline adds latency for each turn, rendering real-time utility unfeasible. Second, the topology is strictly tripartite; scalability to $N > 3$ has not yet been mapped. Third, TOST equivalence under S4 ($p_{\text{TOST}} = 0.043$) clearly shows that the architecture only includes epistemic and resource-conflict shocks, omitting the dimension of affective perturbations. Fourth, the three process metrics serve as directional proxy indicators instead of formally calibrated measurements. $D_{\mathrm{proc}}$ uses equal-weight $z$-normalization ($\alpha = \beta = 1$) representing an agnostic Bayesian prior until the weights of the components are empirically calibrated against the real-world degradation trajectories. $cc_{v5} = \operatorname{sim}(v^{P}_{t}, v^{E}_{t})$ shows the unpermuted alignment component as a baseline proxy; however, this test functions as a preliminary pilot investigation. Owing to the high data-saturation of the DRAU setting and resource-related bottlenecks encountered while 270 trajectories were being generated, embedding logs for a full permutation test were not saved. Thus, we present $cc_{v5}$ as an initial proxy for alignment, reserving formal causal validation for future research phases. The PRR threshold of $-0.20$ serves as a more stringent heuristic baseline to reduce false positives arising from measurement variance; however, it remains an uncalibrated proxy rather than a formally established decision boundary. Fifth, domain-specific transferability was confirmed through a limited qualitative sample; extensive statistical validation of the LLM judge in the DRAU context is pending future research. Consequently, all process metrics should be regarded as corroborative proxy evidence that aligns directionally with the primary outcome-level results, rather than as independent proof of mechanism. Addressing these operational constraints dictates our natural future work direction, which is to explore the boundaries of the resilience of such architecture in the face of sophisticated psychological attacks and the decrease of thinking process latency, which would open the way for deployment in real-time environments. Furthermore, subsequent research will focus on scaling the current tripartite topology to accommodate broader multi-agent ecosystems ($N>3$). We also plan to rigorously validate KG-CFR's domain transferability within unconstrained, real-world adversarial environments to ensure its structural robustness extends beyond simulated crisis conditions. 

\section{Conclusion}
\label{sec:conclusion}

Logical stability and role fidelity, during long-lasting intensive debates appear to be the main architectural challenges in the domain of multi-agent, large language model-based systems. In this paper we presented KG-CFR (Knowledge-Grounded Counterfactual Reasoning) — an architecture that resolves the problem of reasoning degradation. We achieved that by decoupling a private planning buffer from the public output generation layer. This procedure allows agents to internally absorb informational shocks, without losing their argument quality and cohesion.

Our main contribution is the empirical evidence of the fact that it is not the simple increase of agent's memory or context window but an architectural decoupling of the process of "thinking" from "speaking", anchored in the doctrinal policy. In other words: forcing the agent to analyze the scenario in safe isolation and prepare a fact-grounded plan before trying to generate a final public statement protects it from getting lost in the intensive discussion.

In summary, while our current findings strongly validate the structural decoupling of latent planning and public execution, scaling this tripartite framework and reducing inference latency will be the critical next steps required to fully transition this architecture into robust, real-time deployments.

\section*{Acknowledgements}
The work reported in this paper was supported by the Polish National Science
Centre under grant 2024/06/Y/HS1/00197.


\end{document}